\documentclass[preprintnumbers,superscriptaddress,showkeys,byrevtex]{revtex4}
\usepackage{amsmath,amsfonts,amssymb,amscd,amsxtra,amsthm}
\usepackage{graphicx}
\usepackage{epstopdf}
\usepackage{bm}
\usepackage{orcidlink}
\begin{document}
\preprint{PKNU-NuHaTh-2024}
\title{Pion-nucleus elastic scatterings incorporating medium effects\\ within the Eikonal-Glauber model}
\author{Hyeon-dong Han\,\orcidlink{0000-0001-8645-0982}}
\email[E-mail: ]{monlistor@pukyong.ac.kr}
\affiliation{Department of Physics, Pukyong National University (PKNU), Busan 48513, Korea}
\affiliation{Center for Extreme Nuclear Matters (CENuM), Korea University, Seoul 02841, Korea}
\author{Parada T.~P.~Hutauruk\,\orcidlink{0000-0002-4225-7109}}
\email[E-mail: ]{phutauruk@gmail.com, phutauruk@pknu.ac.kr}
\affiliation{Department of Physics, Pukyong National University (PKNU), Busan 48513, Korea}
\author{Seung-il Nam\,\orcidlink{0000-0001-9603-9775}}
\email[E-mail: ]{sinam@pknu.ac.kr,\,\,gariwulf@gmail.com}
\affiliation{Department of Physics, Pukyong National University (PKNU), Busan 48513, Korea}
\affiliation{Center for Extreme Nuclear Matters (CENuM), Korea University, Seoul 02841, Korea}
\affiliation{Asia Pacific Center for Theoretical Physics (APCTP), Pohang 37673, Korea}
\date{\today}
\begin{abstract}
In this present investigation, we explore the elastic scattering of pions with nuclei ($\pi$-$A$), primarily influenced by the $\Delta$(1232) resonance, within the Eikonal-Glauber model. The medium effects are incorporated by considering nuclear-density ($\rho_A$) dependent masses of baryons and strong coupling constants. These dependencies are computed and parameterized up to $\mathcal{O}(\rho_A^2)$ based on the quark-meson coupling (QMC) model. The Wood-Saxon type density profile is utilized for the bound nucleons within finite nuclei. The element $\pi^+$-$N$ scattering cross section for the Glauber approach is determined using the conventional effective Lagrangian method. Subsequently, we analyze the total cross sections for elastic scattering with $^4$He and $^{12}$C targets. Our numerical results demonstrate a favorable agreement with JINR data for the $^4$He target, accurately reproducing the total cross-section. However, when considering the $^{12}$C target, deviations of approximately $\lesssim10\%$. We also consider the multiple-scattering effects inside the nucleus approximately, using the single-channel meson-baryon Bethe-Salpeter equation, resulting in the effective width broadening of the $\Delta$ resonance to reproduce the data better.  
\end{abstract}
\maketitle
\section{Introduction} \label{sec:intro}
The investigation of low-energy pion-nucleus ($\pi$-$A$) scattering has been one of the most important subjects within the nuclear physics community since 1950~\cite{Serber:1947zza,Fernbach:1949zz}. This field holds significance as it offers valuable insights into nucleon-nucleon ($N$-$N$)  interactions and nuclear structure. The $\pi$-$A$ interactions manifest through elastic scatterings, where the target nucleus remains in its ground state~\cite{Landau:1977th,Khankhasayev:1989mp,Aleksakhin:1998qh}, inelastic scatterings with the nucleus in excited states upon pion emission~\cite{Lee:2002eq}, and absorption processes where no pion emerges from the nucleus~\cite{Weyer:1990ye,Golak:2022tjw}. Numerous theoretical studies on $\pi$-$N$ and $\pi$-$A$ scatterings have been conducted using diverse approaches, including the optical model in coordinate space~\cite{Carr:1982zz,Seki:1983sh}, the optical model in momentum space~\cite{Landau:1977th}, coupled channel theory~\cite{Koltun:1980cn}, extended schematic models~\cite{Hirata:1977is}, cloudy bag models~\cite{Jennings:1984ej}, and chiral bag models incorporating quark degrees of freedom~\cite{Thomas:1982kv}. It is noteworthy that theoretical model calculations often account for pion absorption by employing an effective potential, typically proportional to the second order of the nuclear density, denoted as $\mathcal{O}(\rho_A^2)$. 

This is akin to the absorption coefficient, signifying that the rate of particle absorption per unit time at a specific position in the nucleus is proportionate to the particle density at that location. This coefficient is a component of the Schr\"odinger equation within the optical potential, manifesting as an imaginary potential with absorptive characteristics, while the real potential is recognized as dispersive. Alternatively, this absorption coefficient can be derived from the continuity equation, as detailed in Ref.~\cite{Bethe:1940zza}. Nevertheless, contemporary interest in elastic  $\pi$-$A$ scattering has waned due to the intricate nature of nuclear structure. Despite extensive theoretical endeavors, numerous experiments have been conducted, amassing a wealth of data on $\pi$-$A$ scattering cross-sections. However, these experimental findings still bear substantial uncertainties and lack precision owing to limited information about nuclear structure. Moreover, the available experimental data pertains only to lower pion energies, approximately around $70$ MeV. Fortunately, recently, the maximum pion energy has surged to approximately $350$ MeV for $^{12}$C and 208 MeV for $^4$He, respectively, for instance~\cite{Buchle:1989hb}.

In this study, we conduct a phenomenological examination of $\pi$-$A$ elastic scattering, emphasizing the dominance of the $\Delta(1232)$ resonance and incorporating medium effects within the Eikonal-Glauber model (EGM)l, utilizing a $\pi^+$ beam. The medium effects are introduced through nuclear-density ($\rho_A$) dependent masses of baryons and strong coupling constants, computed and parameterized based on the quark-meson coupling (QMC) model~\cite{Guichon:1987jp,Saito:2005rv,Hutauruk:2018qku,Hutauruk:2019was,Hutauruk:2019ipp,Stone:2016qmi} up to $\mathcal{O}(\rho_A^2)$. QMC computations reveal that effective nucleon masses for $A=(^4\mathrm{He},^{12}\mathrm{C})$ gradually increase up to $r \simeq 2$ fm and then become saturated. A similar trend is observed for the $\Delta$ mass, albeit with differing magnitudes. For the calculation of $\pi^+$-$A$ elastic scattering cross-sections, we utilize the element $\pi^+$-$N$ elastic scattering cross-sections computed via the effective Lagrangian approach at the tree-level Bonn approximation. It is essential to highlight that $\pi^+$-$N$ scattering cross-sections are predominantly influenced by the $\Delta(1232)$ resonance, and our numerical results effectively reproduce existing experimental data, as demonstrated in previous research~\cite{Han:2021zrh}.

Our findings indicate that the total cross-section for the $^4$He target closely agrees with the data from the Joint Institute for Nuclear Research (JINR)~\cite{Shcherbakov:1975gm}. However, for the $^{12}$C target, the results tend to overestimate the data by less than $10\%$, particularly when compared to those of the Shanghai Institute of Nuclear Research (SINR)~\cite{Ashery:1981tq}, attributed to the oversimplification inherent in the current Glauber model for heavier nuclei. To improve the theoretical results, the multiple-scattering effects inside the nucleus are taken into account approximately, using the single-channel meson-baryon Bethe-Salpeter equation. As a result, it is observed that the effective broadening of the decay width of the $\Delta$ resonance occurs, resulting in the data are described better in comparison to that without the multiple-scattering effects. We also provide the prediction for the $^{16}$O target case with the same theoretical framework.  
 
This manuscript is structured as follows. Section II provides a brief introduction to the theoretical framework for the $\pi^+$-$A$ elastic scatterings. In Section III, we present comprehensive numerical results encompassing various theoretical aspects, such as effective baryon mass, strong couplings, cross sections, and multiple scatterings for nuclei targets mainly, accompanied by detailed discussions. The concluding section is dedicated to summarizing the key findings.

\section{Theoretical framework and numerical results}
In this section, we provide a brief overview of the current theoretical framework governing elastic $\pi^+$-$A$ scatterings within the EGM. We outline the incorporation of medium effects into the model, specifically addressing the determination of effective baryon masses and relevant strong couplings. These medium effects are computed using the Quark-Meson Coupling (QMC) model up to $\mathcal{O}(\rho_A^2)$, with $\rho_A$ representing nuclear density. Additionally, we account for the bound-nucleon density profile within the nuclei using the Wood-Saxon functional form, given in Ref.~\cite{Woods:1954zz}. Historically, the Glauber model was initially proposed in Refs.~\cite{glauber1970,Glauber1959}, proving highly successful as a phenomenological model for characterizing nuclear properties, particularly in $N$-$A$ and $A$-$A$ interactions observed in heavy-ion collision (HIC) experiments, as documented in Ref.~\cite{Miller:2007ri}. This model interprets $N$-$A$ scattering in terms of multiple elastic $N$-$N$ scatterings. Our focus in this study is on the total cross-sections of elastic $\pi$-$A$ scatterings involving different nucleus targets, such as $^4$He and $^{12}$C. To this end, we employ the EGM, an updated version of the Glauber approach that assumes a projectile follows a linear path, known as the Eikonal approximation~\cite{Goldstein:2006sv}.

It is crucial to emphasize that, in our computations employing the current EGM, we rely on the element $\pi$-$N$ scattering cross-sections as input, where the dominant influence is attributed to the $\Delta(1232)$ resonances. Within the framework, the expression for the total cross-section is defined as follows:
\begin{eqnarray}
  \label{eqN15}
  \sigma_{\pi A} (s^*) = \int d^2\mathbf{b}  \Big[1-\exp\left[-\tilde{\sigma}_{\pi N} (E^*_\mathrm{cm},\rho^N_A)\,T^N_A(\mathbf{b})\right] \Big],\,\,\,\,
   T^N_A(\mathbf{b}) = \int^{\infty}_{-\infty} dz\, \rho^{N}_{\mathrm{A}} \left(\sqrt{|\mathbf{b}|^2+z^2}\right),
\end{eqnarray}
Here, $\tilde{\sigma}_{\pi N}$ represents the sum of the $\pi$-$N$ scattering cross-section for $N=(n,p)$ within the nucleus, and $T^N_A$ denotes the thickness function describing the transverse reaction probability at the two-dimensional impact-parameter vector $\mathbf{b}$. Meanwhile, $z$ signifies the longitudinal spatial position. The combined cross-section, accounting for the numbers of neutrons ($n$) and protons ($p$) within the nuclei, can be expressed as follows:
\begin{eqnarray}
  \label{eqN16}
  \tilde{\sigma}_{\pi N}=\left(\frac{A-Z}{A}\right) 
  \sigma_{\pi n}+\left(\frac{Z}{A}\right) \sigma_{\pi p}.
\end{eqnarray}
It is worth noting that the ratio between the elementary $\pi$-$N$ cross sections, namely $\sigma_{\pi p}$ and $\sigma_{\pi n}$, is approximately 9:1. This ratio can be understood by their distinct isospin factors in elastic scatterings, as elucidated in prior research~\cite{Han:2021zrh}.

\begin{figure}[t]
\begin{tabular}{cc}
 \includegraphics[width=8cm]{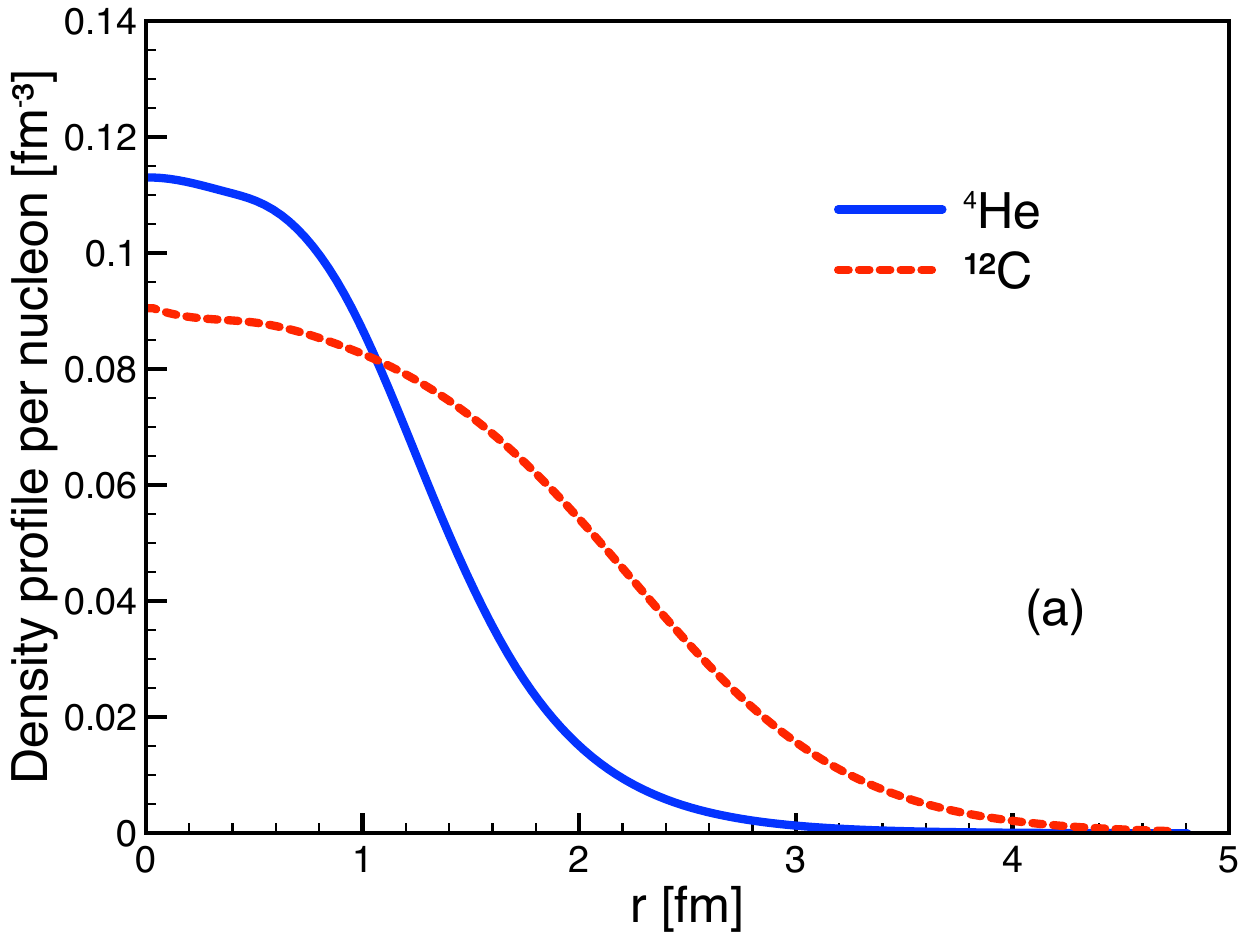}
 \includegraphics[width=8cm]{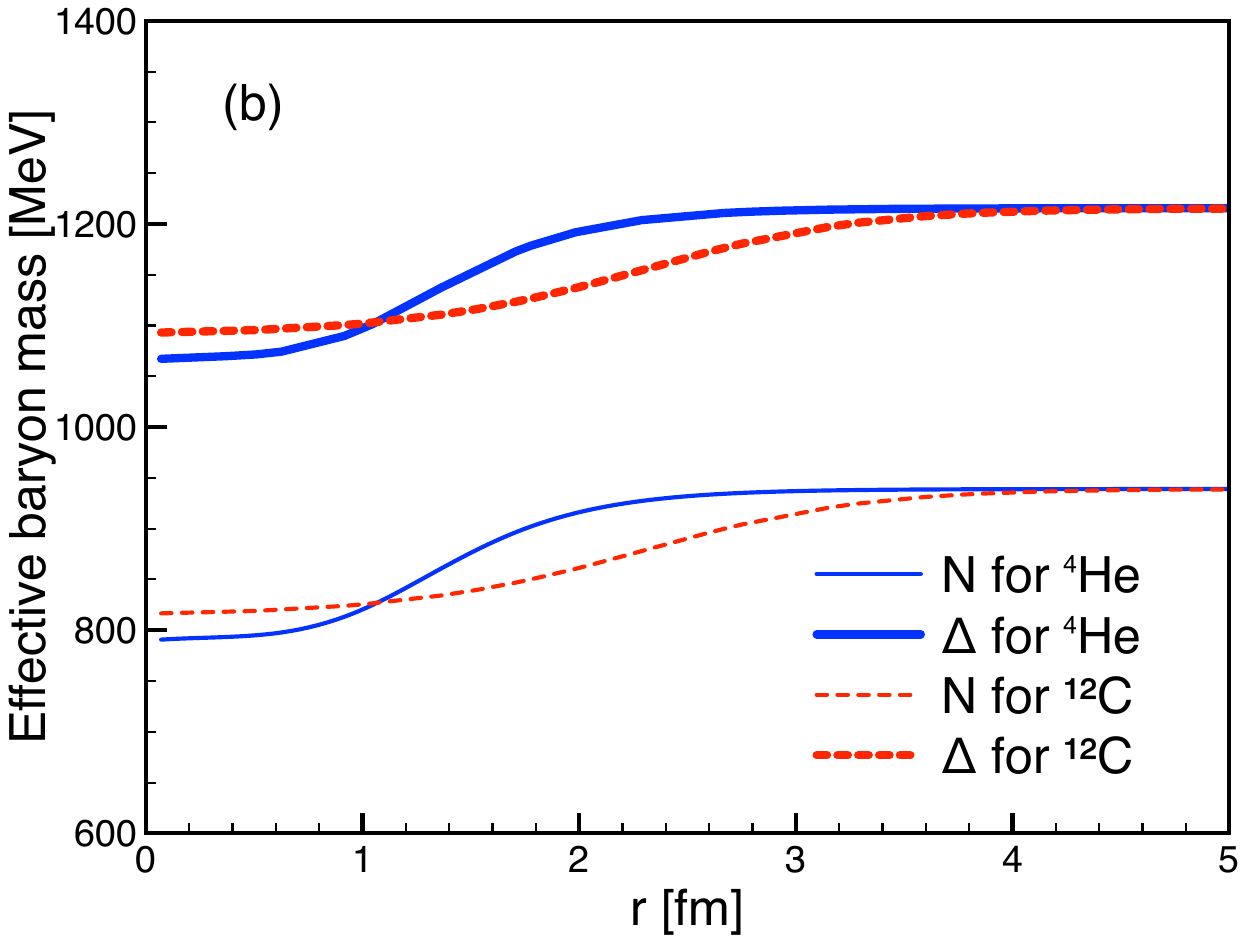}
 \end{tabular}
\caption{(a Density profile per nucleon $\rho^{N}_A$ [fm$^{-3}$] as a function of the radial distance $r$ [fm] for $^4$He (solid) for $^{12}$C (dashed). (b) Effective baryon mass $M^*_B$ [MeV] for $B = N,\Delta$ in the same manner with the panel (a).}
\label{FIG12}
\end{figure}

To consider the internal density distribution of the nuclei, the Wood-Saxon density profile, presented as a function in Eq.~(\ref{eqN15}), is expressed as follows:
\begin{eqnarray}
  \label{eqN17}
\rho^{(n,p)}_{\mathrm{A}} (r) = \left[ \frac{\rho^c_A \left[ 1+c(r/r_A)^2 \right]}{1 + \mathrm{exp}[(r-r_A)/d]} \right]
\left[\frac{(A-Z,Z)}{A} \right],\,\,\,\,r = \sqrt{|\mathbf{b}|^2+z^2},
\end{eqnarray}
Here, $r$, $\rho^c_A$, $r_A$, $d$, and $c$ represent the radial distance, center density, average radius, surface thickness, and deformation parameter for the target nucleus, respectively. The specific values for these parameters are determined through fitting to experimental data, as detailed in Ref.~\cite{DeJager:1974liz,DeVries:1987atn}, and are compiled for $^4$He, $^{12}$C, and $^{16}$O in Table~\ref{table1}.
\begin{table}[b]
\begin{tabular}{ c | c | c | c | c | c | c } \hline
Nuclei& $A$ & $Z$ & $r_A$ (fm) & $d$ (fm) & $\rho^c_A$ (fm$^{-3}$) & $c$ \\ \hline
$^{4}$He & 4 & 2 & 1.01 & 0.327 & 0.2381 & 0.445 \\ 
$^{12}$C & 12 & 6 & 2.36 & 0.522 & 0.1823 & $-$0.149 \\
$^{16}$O & 16 & 8 & 2.608 & 0.513 & 0.1701 & $-$0.051 \\ \hline
\end{tabular}
\caption{$r_A$, $d$, $\rho^c_A$, and $c$ for $^4$He, $^{12}$C, and $^{16}$O from the fitting of experimental data~\cite{DeJager:1974liz,DeVries:1987atn}. }
  \label{table1}
\end{table}
The density profile in Eq.~(\ref{eqN17}) satisfies the following normalization condition:
\begin{equation}
\label{eqN18}
\sum_{N=(n,p)}\int dz\,\, d^2\mathbf{b}\,\, \rho^{N}_A (\sqrt{|\mathbf{b}|^2+z^2} )= A.
\end{equation}
This implies the conservation of nucleon number density. The density profiles for $^4$He (solid) and $^{12}$C (dot-dashed) are depicted in panel (a) of Fig.~\ref{FIG12}, showcasing their dependence on the radial distance $r$ from the center. Given that we are specifically examining even-even nuclei, it follows that $\rho^n_A=\rho^p_A=\rho_A$.

As previously mentioned, in this study, we determine the density-dependent effective masses for baryons, along with the corresponding strong coupling constants, utilizing the QMC model within the mean-field approximation. Further details regarding these calculations can be found in the prior work~\cite{Han:2021zrh} and the references therein. For practical purposes, we parameterize these quantities as functions of $\rho_A$ up to $\mathcal{O}(\rho^2_A)$, akin to the imaginary potential for $A$-$A$ derived in the Schr\"odinger equation. Consequently, the parameterization for the effective baryon mass $M^*_B$ is expressed as follows:
\begin{eqnarray}
\label{eqN19}
M^*_B &=& M_B + C_1\,\rho_A+ C_2\,\rho^2_A +\mathcal{O}(\rho^{n>2}_A),
\end{eqnarray}
where $M_B$ stands for the vacuum mass, and the relevant coefficients are given by $C_1 =-1.543\,\mathrm{GeV\cdot fm}^3$ and  $C_2= 2.036\,\mathrm{MeV\,fm}^3$ via the QMC calculation. 

The panel (b) of Fig.~\ref{FIG12} illustrates the variation of $M^*_B$ for $^4$He and $^{12}$C with respect to the radial distance $r$. In the QMC model, the effective nucleon masses $M^*_N$ exhibit an initial value of approximately $800$ MeV at the nuclei's center due to the partial restoration of spontaneous chiral symmetry breakdown (SCSB) at finite densities. As one moves radially outward, the medium effects diminish, leading to a convergence toward the nucleon vacuum mass of approximately $\sim940$ MeV. It is noteworthy that the behavior of $M^*_\Delta$ for the nuclei mirrors that of the nucleon, given the utilization of the same mass formula in Eq.~(\ref{eqN19}) for both baryons. However, the $M^*_{N,\Delta}$ curves display distinct dependencies on $r$ for different nuclei. This discrepancy arises from their disparate density profiles, as depicted in panel (a) of Fig.~\ref{FIG12}. In addition to examining the influence of the medium effects on masses, we also investigate alterations in the strong coupling constants. To address this, we incorporate the Goldberger-Treiman relation (GTR) concerning the weak axial-vector coupling constant $f_A$, which is expressed as follows:
\begin{eqnarray}
f_A &=& \frac{f_{\pi NN}\, f_\pi}{M_N}.
\end{eqnarray}

The numerical outcomes for $M^*_B$ as functions of $r$ are presented in panel (b) of Fig.~\ref{FIG12} for $^4$He and $^{12}$C. The effective nucleon masses $M_N$ commence at approximately $800$ MeV at the nucleus's center, reflecting the partial restoration of the spontaneous breakdown of chiral symmetry (SBCS) at finite densities within the QMC model. With increasing radial distance, medium effects diminish, converging to the nucleon vacuum mass of $\sim940$ MeV. Notably, $M^*_\Delta$ for the nuclei exhibits similar behaviors to those of the nucleon, given the employment of the same mass formula in Eq.~(\ref{eqN19}) for both baryons. However, the $M^*_{N,\Delta}$ curves display distinct $r$ dependencies contingent on the nuclei, stemming from their diverse density profiles as illustrated in panel (a) of Fig.~\ref{FIG12}. 
\begin{equation}
\label{eq:FFFF}
f^*_{\pi NN} (\rho_A) =f_{\pi NN} - 0.633\,\rho_A,\,\,\,\,
 f^*_{\pi N \Delta}(\rho_A) = f_{\pi N\Delta} - 1.360\,\rho_A.
\end{equation}
Here, we adopt the values $f_{\pi NN} = 0.989$, $f_{\pi N \Delta} = 2.127$, and $f_\pi = 93.2$ MeV, sourced from the Nijmegen potential model~\cite{Gasparyan:2003fp} and experimental data~\cite{Janssen:1996kx}. The determination of $f^*_{\pi N \Delta}$ assumes that the ratio $f_{\pi NN}/f_{\pi N\Delta} \approx 2.15$ remains constant within the medium.

\begin{figure}[t]
\begin{tabular}{cc}
\includegraphics[width=8cm]{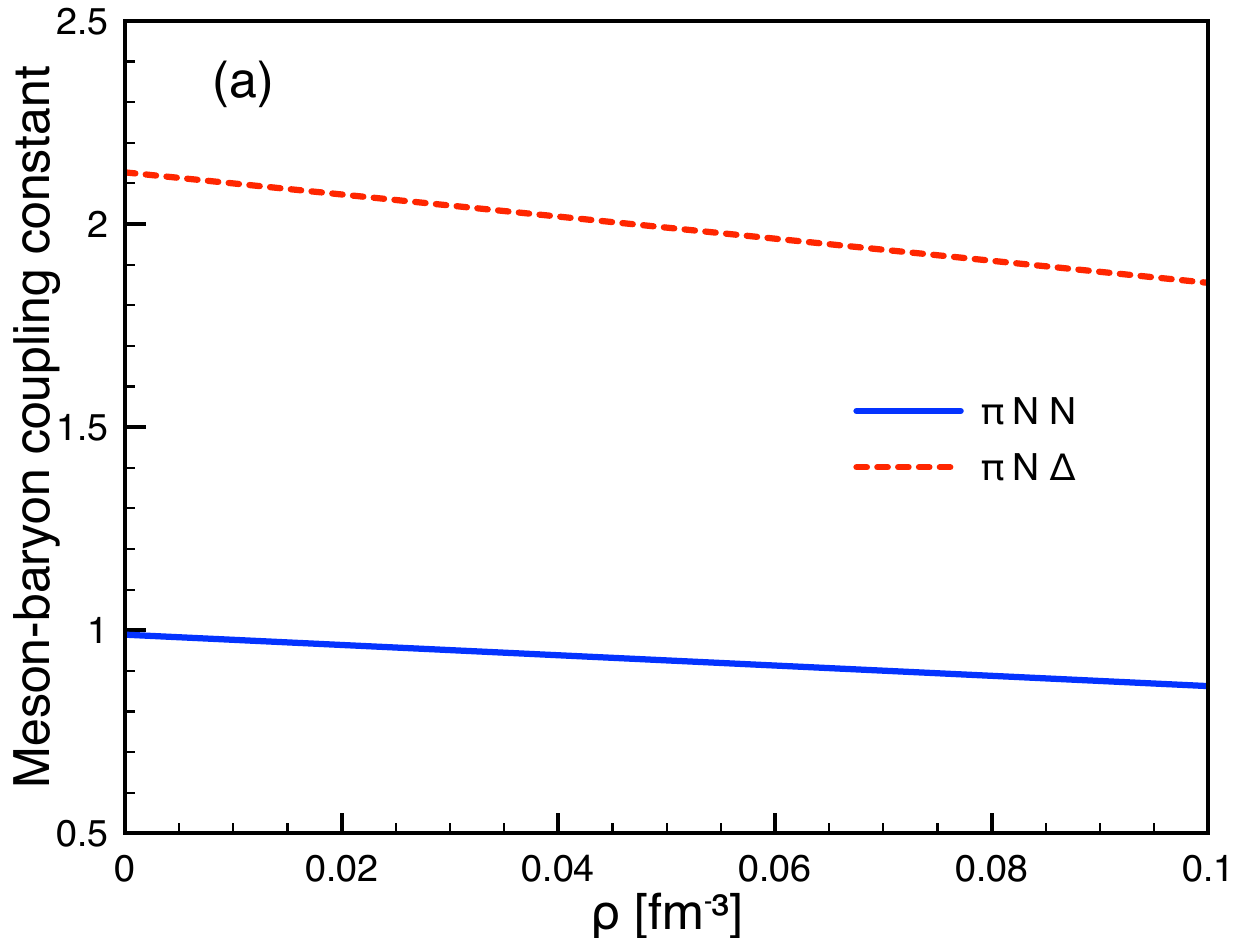}
\includegraphics[width=8.5cm]{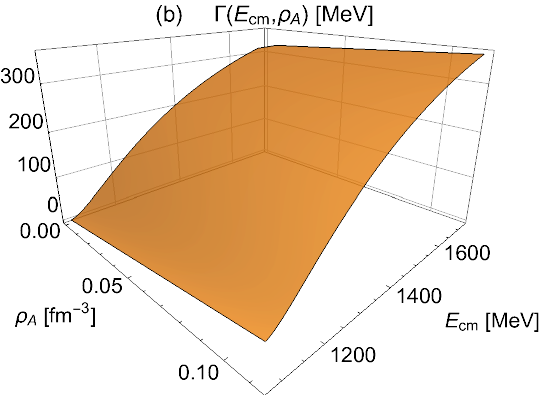}
\end{tabular}
\caption{(a) Meson-baryon coupling constants $f_{\pi N N}$ (solid) and $f_{\pi N\Delta}$ (dashed) as functions of $\rho_A$ [fm$^{-3}$]. (b) $\Delta$-baryon decay width in Eq.~(\ref{eqN20}), $\Gamma(E_\mathrm{cm},\rho_A)$ [MeV] as a function of $E_\mathrm{cm}$ [MeV] and $\rho_A$ [fm$^{-3}$].}
\label{FIG34}
\end{figure}
Our findings for the in-medium modifications of the pion-nucleon coupling constants $f^*_{\pi NN}$, the pion-delta coupling constant $f^*_{\pi N \Delta}$, and the pion decay constant with medium correction effects, as functions of $\rho_A$, are illustrated in Fig.~\ref{FIG34}. In panel (a) of Fig.~\ref{FIG34}, it is evident that the density dependence of the pion decay constant diminishes with increasing $\rho_A$, aligning with results from other theoretical models. Similar trends are observed for $f_{\pi NN}^*$ and $f_{\pi N\Delta}^*$, both decreasing as $\rho_A$ increases, albeit at a slower rate compared to $f_\pi^*$. It is important to note that these medium corrections will be utilized as input when calculating the cross-section of pion-nucleus scattering for $^4$He and $^{12}$C.

In our calculations, we also account for the momentum-dependent $\Delta$ decay width $\Gamma$, as employed in Ref.~\cite{Larionov:2003av}. However, in finite nuclei, the decay width should be determined by the nuclear density distribution $\rho_A$, taking the form:
\begin{equation}
\label{eqN20}
 \Gamma(E^*_\mathrm{cm},\rho_A)= \Gamma_{\mathrm{sp}} \left(\frac{\rho_A}{\rho_0} \right) + \Gamma_0\left[\frac{q\left(M^*_N,M_\pi,E^*_\mathrm{cm}\right)}
{q\left(M_N,M_\pi,M_\Delta\right)}\right]^3  \left[\frac{M^*_\Delta}{E^*_\mathrm{cm}}\frac{\beta^2_0+q^2\left(M^*_N,M_\pi,M^*_\Delta\right)}{\beta^2_0+q^2\left(M^*_N,M_\pi,E^*_\mathrm{cm}\right)} \right],
\end{equation}
where $\Gamma_{\mathrm{sp}} = 80$ MeV represents the spreading density dependence of the decay width, determined through fitting to the medium quantity at finite density. Additionally, $\beta_0 = 200$ MeV serves as the cutoff parameter, consistent with the choice made in Ref.~\cite{Larionov:2003av}. The results for the 3D plot of the decay width for the $\Delta$ baryon in finite nuclei, as a function of $\rho_A$ and $E^*_\mathrm{cm}$, are depicted in Fig.~\ref{FIG34}. This visualization indicates that the in-medium modifications of the $\Delta$ baryon decay width differ significantly for $^4$He and $^{12}$C due to their distinct nuclear charge density distributions of $\rho_A$, as illustrated in Fig.~\ref{FIG34}. The in-medium modifications of the $\Delta$ baryon decay width increase with rising nuclear density distribution $\rho_A$. Furthermore, it is observed that as $E_\mathrm{cm}$ increases, the in-medium modifications of the $\Delta$ baryon resonance decay width $\Gamma^*_\Delta$ also increase, consistent with the findings in the calculation of Ref.~\cite{Cui:2020fhr}.

\begin{figure}[t]
\begin{tabular}{cc}
\includegraphics[width=8cm]{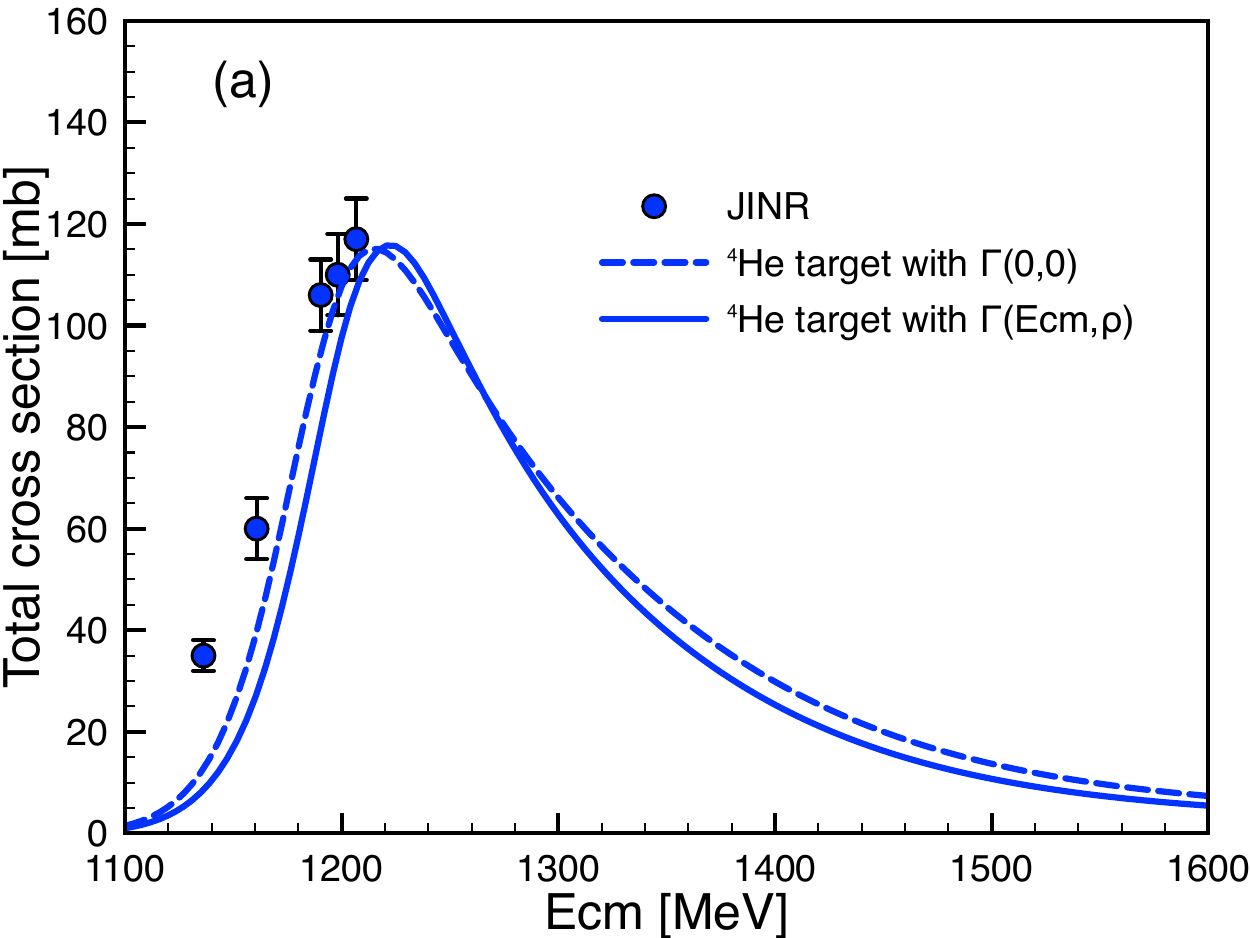}
\includegraphics[width=8cm]{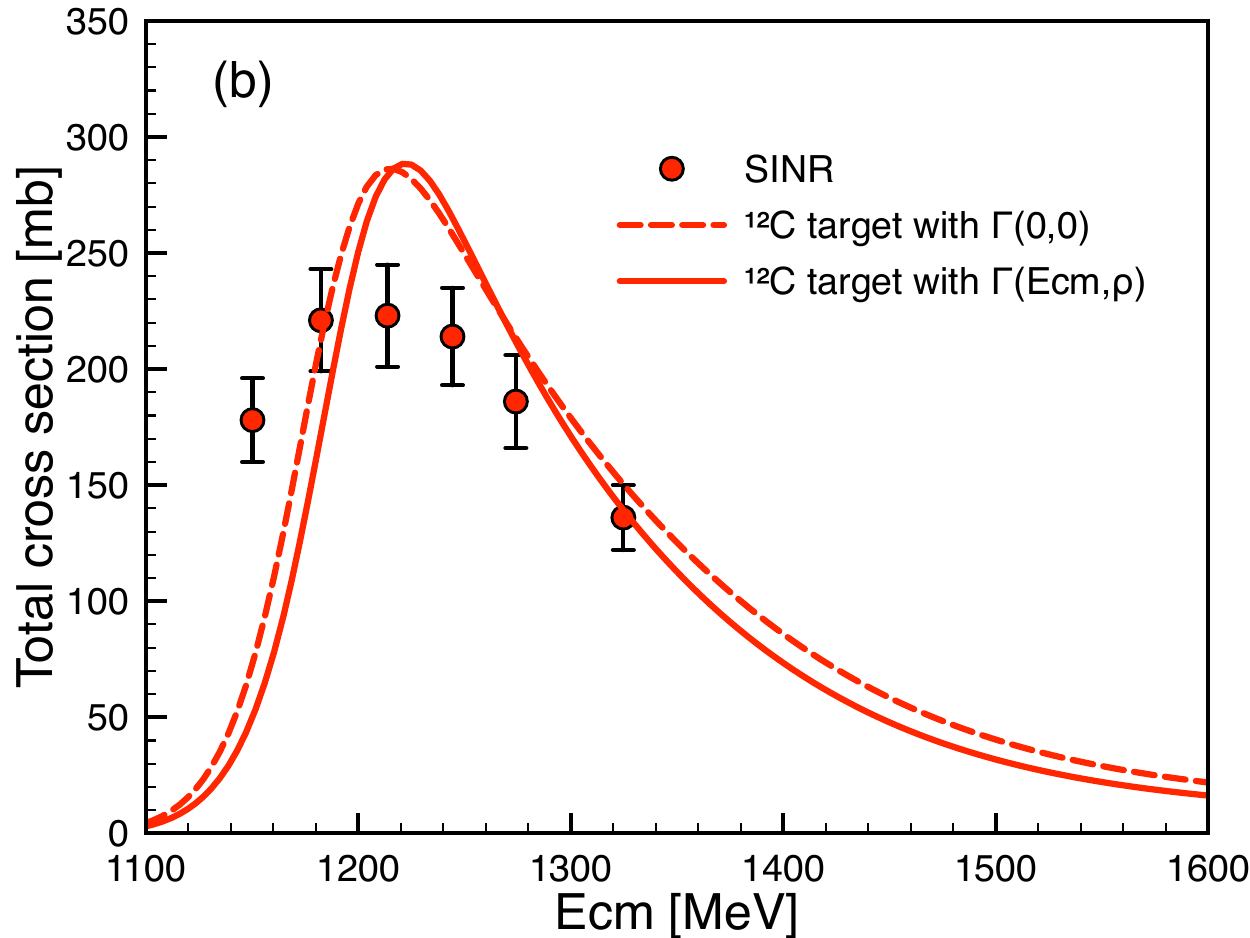}
\end{tabular}
\caption{(a) Total cross-section [mb] for the $^4$He target as a function of the center-of-mass (cm) energy $E_\mathrm{cm}$ [MeV]. Here, we consider two cases for the $\Delta$ decay width in Eq.~(\ref{eqN20}) with (dashed) and without (dashed) the $(E_\mathrm{cm},\rho_A)$ dependence. The experimental data are taken from JINR.~\cite{Shcherbakov:1975gm}. (b) The same for the $^{12}$C target. The data are taken from SINR~\cite{Ashery:1981tq}.}
\label{FIG56}
\end{figure}

Now, we are in a position to present the numerical results for the cross-section of elastic pion-nucleus scattering at intermediate energies, where $\Delta$ baryon resonance dominates, for both $^4$He and $^{12}$C in Fig.~\ref{FIG56}. For a more in-depth exploration of the elastic pion-proton and pion-neutron scatterings, interested readers are referred to Ref.~\cite{Han:2021zrh}. The numerical results for the elastic cross-sections for the $^4$He target, accounting for medium corrections up to $\rho_A^2$ with $\tilde{\Gamma}^*_\Delta(E_\mathrm{cm},\rho_A)$ as a function of center-of-mass (cm) energy, are presented in the panel (a) of Fig.~\ref{FIG56}, alongside JINR experimental data\cite{Shcherbakov:1975gm}. Fig.~\ref{FIG56} clearly illustrates the total cross-section for $^4$He, being qualitatively consistent with the  data, particularly around the resonance region at $E_{\mathrm{cm}} \sim$ 1222 MeV. The dashed and solid lines denote the results with and without the energy- and density-dependent decay width, respectively, not showing considerable difference. For the $^{12}$C, the numerical result, compared with SINR experimental data~\cite{Ashery:1981tq}, is shown in the panel (b) of Fig.~\ref{FIG56}. Here, the the numerical result overestimates the data\cite{Ashery:1981tq} by about $10\%$ percent deviations around the resonance peak, in contrast to the result for $^4$He. Again, the modified width in Eq.~(\ref{eqN20}) does not make sizable difference even for the $^{12}$C target. In what follows, we would like to discuss the deviation between the theory and experiment for the $^{12}$C target in detail. 

\begin{figure}[t]
\begin{tabular}{cc}
\includegraphics[width=8cm]{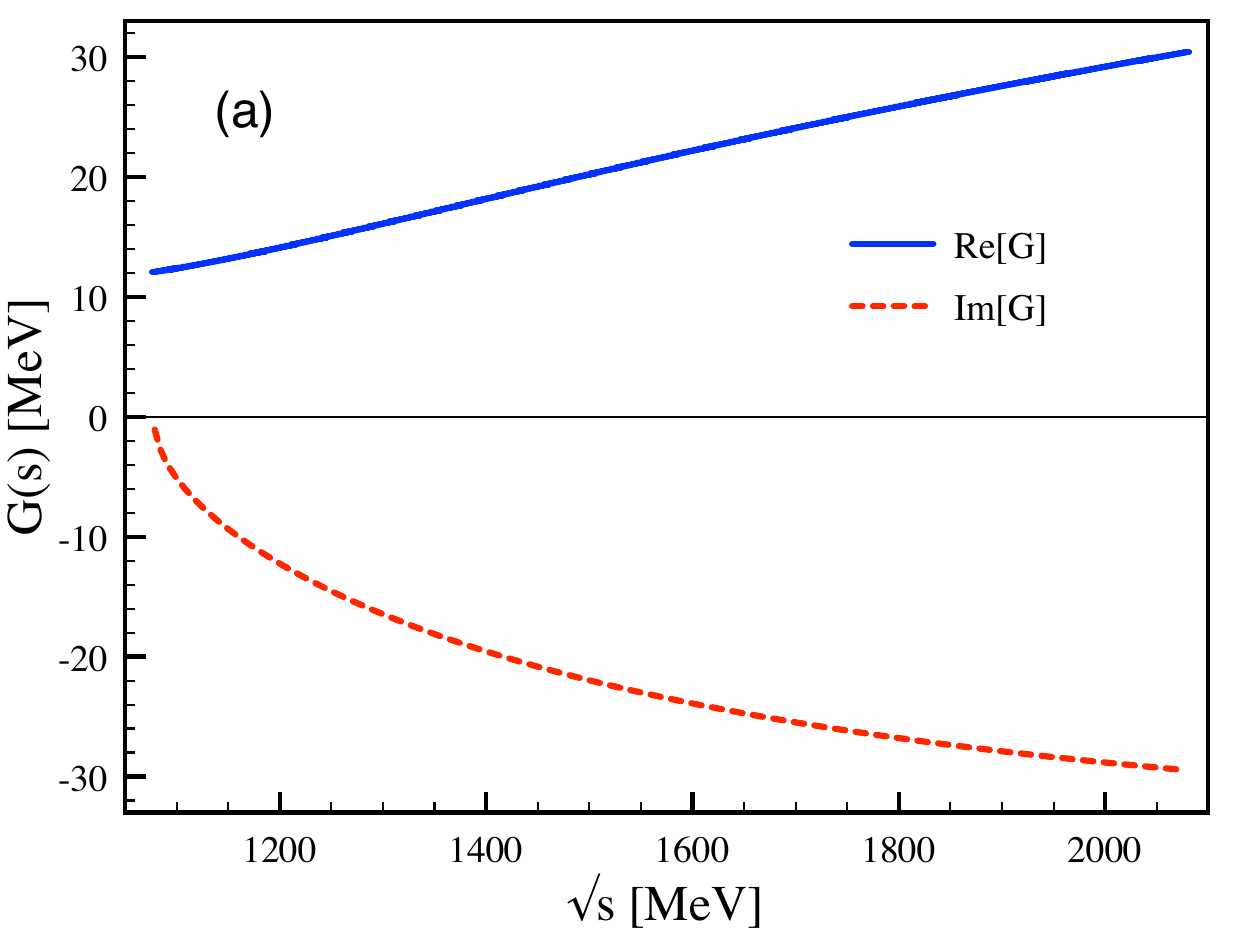}
\includegraphics[width=8cm]{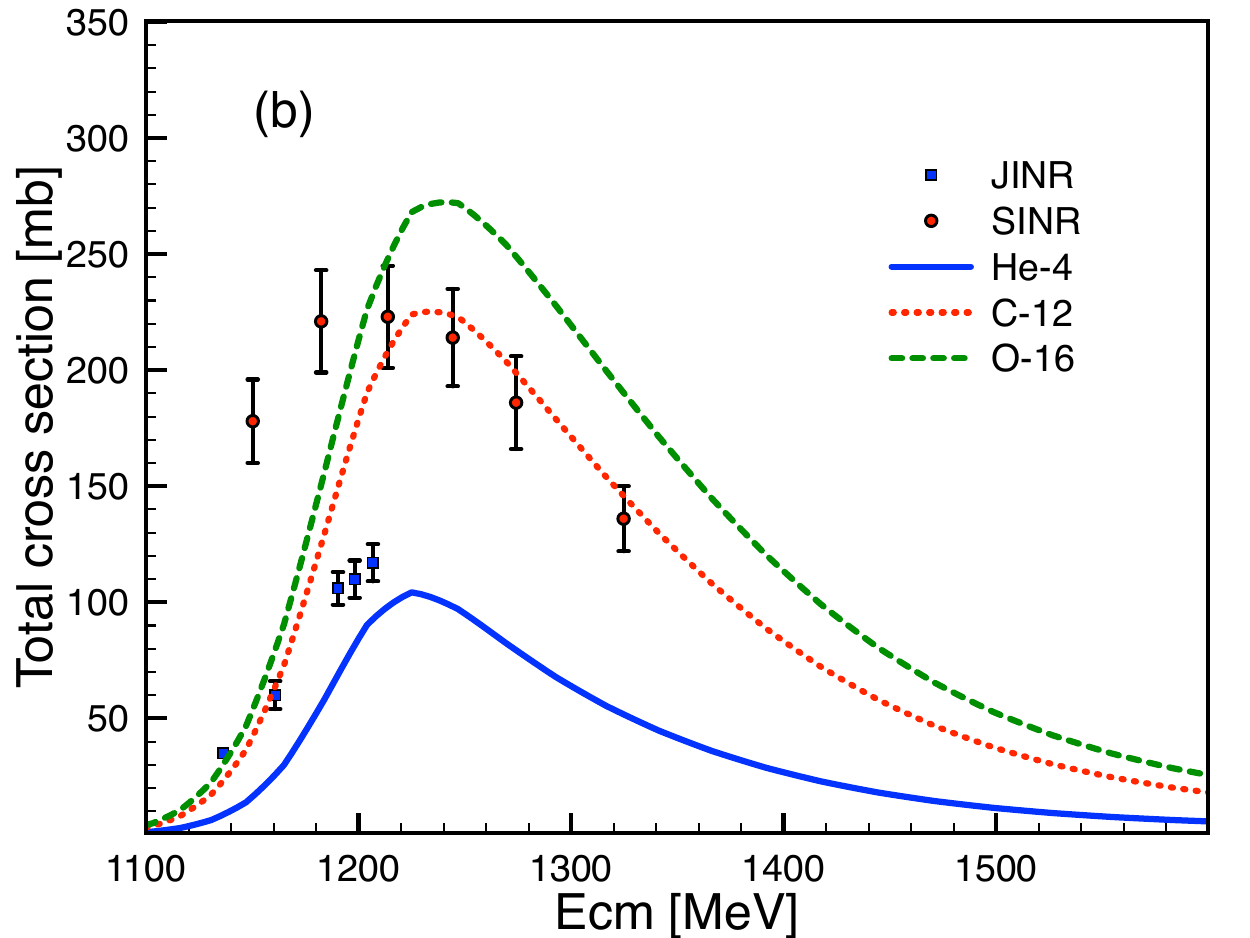}
\end{tabular}
\caption{(a) Real (solid) and imaginary (dashed) parts for the propagator $G_{\pi N}(s)$ in Eq.~(\ref{eq:GFUN}) beyond the $\pi$-$N$ threshold. (b) Numerical results for the elastic cross sections for the  $^{4}$He (solid), $^{12}$C (dotted), and $^{16}$O (dashed) targets as functions of $E_\mathrm{cm}$. The data are taken from Refs.~\cite{Shcherbakov:1975gm,Ashery:1981tq}.}
\label{FIG78}
\end{figure}

In principle, the decay width of the resonance can be modified by the multiple scattering (MS) inside the nucleus, and this modification can be the cause to explain the deviation observed above. Hence, we would like to estimate the effects of the meson-baryon MS inside the nucleus using the single-channel ($\pi$-$N$) Bethe-Salpeter equation, resulting in the broadening decay width of the $\Delta$ resonance, which dominates the reaction process. The Bethe-Salpeter equation for the scattering can be defined as follows:
\begin{equation}
\label{eq:ASAAA}
\tilde{\mathcal{M}}^\mathrm{MS}_{ij}\approx\tilde{\mathcal{M}}_{ij}+\sum_k\tilde{\mathcal{M}}_{ik} G_{k}\tilde{\mathcal{M}}_{kj}+\cdots=\tilde{\mathcal{M}}_{ij}+\sum_k\tilde{\mathcal{M}}_{ik} G_{k}\tilde{\mathcal{M}}^\mathrm{MS}_{kj},
\end{equation}
where $\tilde{\mathcal{M}}$ [mass$^{-1}$] indicates the reduced amplitude for an element process with the channel indices $i,j,\cdots$. It relates to the invariant amplitude by
\begin{equation}
\label{eq:AMAMAM}
\bar{u}_i\tilde{\mathcal{M}}_{ij}u_j=\mathcal{M}_{ij}\,[\mathrm{mass}^{2}].
\end{equation}
Here, $G_k=G_k(s=E^2_\mathrm{cm})$ stands for the meson-baryon propagator in the on-mass-shell approximation for the intermediate $k$ channel. In the present work, for simplicity, we only consider the $\pi N$ elastic channel as mentioned previously, i.e., $(i,j)=(\pi N,\pi N)$. Then, the propagator is given by the dimensional regularization~\cite{Inoue:2001ip} as follows:
\begin{eqnarray}
\label{eq:GFUN}
G_{\pi N}(s)&=&
\frac{M_N}{8\pi^2}\Big[a+\ln\frac{M^2_N}{\mu^2}+
\frac{s-s'_{\pi N}}{2s}\ln\frac{m^2_\pi}{M^2_N}
\cr
&+&\frac{s_{\pi N}}{2s}
\ln\left[(s_{\pi N}+s'_{\pi N}+s)+(s_{\pi N}-s'_{\pi N}+s)-(s_{\pi N}+s'_{\pi N}-s)-(s_{\pi N}-s'_{\pi N}-s)\right]
\Big]\,[\mathrm{mass}],
\end{eqnarray}
where $s_{\pi N}$ and $s'_{\pi N}$ are defined by
\begin{equation}
\label{eq:SSSSS}
s_{\pi N}=s_{\pi N}(s)=\sqrt{[s-(M_N-m_\pi)^2][s-(M_N+m_\pi)^2]},
\,\,\,\,s'_{\pi N}=M^2_N-m^2_\pi.
\end{equation}
As for the $\pi$-$N$ elastic-scattering channel, the renormalization scale $\mu$ and subtraction parameter $a$ are determined by $1.2$ GeV and $2$, respectively, to reproduce available data for the elementary cross-sections~\cite{Inoue:2001ip,Nam:2003ch}. In the panel (a) of Fig.~\ref{FIG78}, we draw the real (solid) and imaginary (dashed) parts of the $\pi$-$N$ propagator in Eq.~(\ref{eq:GFUN}) beyond its threshold. For instance, the single-channel elastic scattering for the $\Delta(1232)$ can be modified by the unitarization of the MS amplitude, due to the multiple scatterings inside the nuclei as follows:
\begin{eqnarray}
\label{eq:MSMS1}
\tilde{\mathcal{M}}^{\Delta,\mathrm{MS}}_{\pi N}&=&\frac{\tilde{\mathcal{M}}^\Delta_{\pi N}}{1-\tilde{\mathcal{M}}^\Delta_{\pi N}G_{\pi N}}
\approx\frac{\hat{\mathcal{M}}^\Delta_{\pi N}}{s-M^2_{\Delta}-i\Gamma_{\Delta}M_{\Delta}}\left(1-G_{\pi N}\frac{\hat{\mathcal{M}}^\Delta_{\pi N}}{s-M^2_{\Delta}-i\Gamma_{\Delta}M_{\Delta}}\right)^{-1},
\end{eqnarray}
where, considering the $\Delta$-resonance dominance, 
\begin{equation}
\label{eq:MSMS2}
\hat{\mathcal{M}}^\Delta_{\pi N}\propto-i\frac{f_{\pi N\Delta}^2}{M_\pi ^2}(k_i\cdot k_f).
\end{equation}
Finally, Eq.~(\ref{eq:MSMS1}) turns into
\begin{eqnarray}
\label{eq:MSMS3}
\tilde{\mathcal{M}}^{\Delta,\mathrm{MS}}_{\pi N}
&=&\frac{\hat{\mathcal{M}}^\Delta_{\pi N}}{s-M^2_{\Delta}
-i\Gamma_{\Delta}M_{\Delta}+i\xi G_{\pi N}\frac{f_{\pi N\Delta}^2}{M_\pi ^2}(k_i\cdot k_f)}.
\end{eqnarray}

As understood by Eq.~(\ref{eq:MSMS3}), the multiple scatterings change the decay width as well as the pole position. It is easy to understand from Eq.~(\ref{eq:ASAAA}) that other contributions besides the resonance also get a small but finite effects from the multiple-scattering mechanism. Note that , in Eq.~(\ref{eq:MSMS3}), we introduce a free parameter $\xi=-15.8\times A$ [MeV] to fit the data. The reasoning for $\xi\propto A$ is that the elementary MS process will be enhanced by the number of the nucleons inside the nucleus. In the panel (b) of Fig.~\ref{FIG78}, we show the numerical results for the elastic cross-sections for the $^{4}$He (solid), $^{12}$C (dotted), and $^{16}$O (dashed) targets, being compared with the data. It is observed that the numerical result is considerably improved for the $^{12}$C, whereas it slightly underestimates the $^{4}$He data. We show the theoretical prediction for the $^{16}$O target as well. As a consequence, it turns out that the width-broadening due to the MS effect is crucial to reproduce the heavier-nucleus data. 

\section{Summary}
In summary, our investigation focused on the elastic $\pi^+$-$A$ scattering process at intermediate energies dominated by the $\Delta(1232)$-resonance, specifically in the $I=$ 3/2 channel, incorporating medium corrections up to $\rho_A^2$ parameterized from the quark-meson coupling (QMC) model within the Eikonal-Glauber model approach. We proceeded to compute the total cross-sections for $^4$He and $^{12}$C targets, utilizing elementary $\pi$-$N$ cross-sections as input, which were calculated within the effective Lagrangian approach at the tree-level Bonn approximation.

For the total cross-section of $^4$He, our results exhibited relative consistency with the JINR data~\cite{Shcherbakov:1975gm}, particularly around the resonance peak. In contrast, that for $^{12}$C was found to overestimate the SINR experimental data~\cite{Ashery:1981tq}. Notably, the resonance peak for $^{12}$C shifted to a lower cm energy, accompanied by a broadened decay width, compared to the corresponding quantities for $^4$He. This observation aligns with findings from other calculations. It was observed that the momentum- and density-dependent decay width does not considerable modifications to the theoretical results. Finally, we took into account the multiple-scattering effect inside the nucleus, using the single-channel Bethe-Salpeter equation, resulting in the with-broadening of the dominating $\Delta$ resonance. We found that the heavier-nucleus cross-section is improved much by the MS effect, describing the $^{12}$C data qualitatively well, whereas the light-nucleus data for the $^{4}$He is reproduced as well with some deviations. 

To properly assess the results of this study and compare them with existing theoretical calculations, it is imperative to acquire new data for the elastic $\pi^+$-$A$ scattering reaction process at intermediate energies in future experiments. Additionally, advancing rigorous theoretical approaches is essential for a clearer understanding of the intricate interactions within both light and heavy nuclei. Related works are in progress and will appear elsewhere.
\section*{Acknowledgements}
This work was supported by a research grant from Pukyong National University (PKNU) (2022).

\end{document}